\documentstyle[12pt]{article}
\begin{document}
%\baselineskip=1.5\baselineskip
\baselineskip=1.2\baselineskip
\renewcommand{\thefootnote}{\fnsymbol{footnote}}
\setcounter{equation}{0}
\setcounter{section}{0}
\renewcommand{\thesection}{\arabic{section}}
\renewcommand{\theequation}{\thesection.\arabic{equation}}
\pagestyle{plain}
\begin{titlepage}

\hfill{SPhT/94-086; ANL-HEP-PR-94-32}
\begin{center}
{\large{\bf  {Higher order corrections to the equation-of-state \\}
{ of QED at high temperature}}}\par
\end{center}
\vskip 1.2cm
\begin{center}
{Rajesh R. Parwani$^{a,}$\footnote{email : parwani@wasa.saclay.cea.fr}
\ and \ Claudio Corian\`{o}$^{b,}$\footnote{email :
coriano@hep.anl.gov}}
\end{center}
\vskip 0.4cm
\centerline{$^a$Service de Physique Th{\'e}orique, CE-Saclay}
\centerline{ 91191 Gif-sur-Yvette, France.}
\vskip 0.3cm
\centerline{$^b$High Energy Physics Division}
\centerline{Argonne National Laboratory}
\centerline{9700 South Cass, Il 60439, USA.}
\vskip 1.0cm
%\centerline{PACS 12.20.Ds, 52.60.+h, 11.15.Bt, 12.38.Mh.}

\centerline{ June 1994; Revised August 1994}

\vskip 1.5 cm
\centerline{\bf Abstract}
{ We elaborate on the computation of the pressure of thermal
 quantum electrodynamics, with massless electrons,
to the fifth ($e^5$) order.
The calculation is performed within the Feynman gauge and the
imaginary-time formalism is employed.
For the $e^4$ calculation, the method of Sudakov decomposition
is used to
evaluate some ultraviolet finite
integrals which have a collinear singularity. For the
$e^5$ contribution,  we give an alternative derivation
and extend the discussion to massive electrons and nonzero
chemical potential.
Comments are made
on expected similarities and differences for
prospective three-loop calculations in QCD.}\\

\end{titlepage}

\newpage
\section{Introduction}

Recently we reported on the order $e^4$ \cite{CoPa} and
$e^5$ \cite{P}  contributions to the pressure of a massless
QED plasma at temperature $T$
and zero chemical potential $\mu_e$. Here we would
like to fill in the
discussion, particularly for the more difficult $e^4$ calculation, as
some of the techniques might be of wider interest. In addition, an
alternative  derivation of the $e^5$ contribution will be given.

For motivation we note that the equation-of-state
of a relativistic plasma is
of relevance in astrophysics \cite{AP,FM,B,H}.
It was obtained at nonzero
$T$ and $\mu_e$ by
Akhiezer and Peletminskii \cite{AP} to the third order
($e^3$) and later
extended to order $e^4$
at $T=0$, but nonzero $\mu_e$, in both QED and
quantum chromodynamics
(QCD)\cite{FM,B}. Thus the calculation of the $T \neq 0, \mu_e=0$
contribution at
order $e^4$ fills a gap in our knowledge. It is also a step towards
same order calculations for QCD which is believed to exist
in a perturbative
quark-gluon phase at high temperature. Currently the pressure of
the high-temperature phase of QCD is known to order
$g^4 \ln g$, excluding the normalisation of the
logarithm \cite{K,T,KKT}.

The calculations in this paper are performed within
the framework of the imaginary-time formalism (see \cite{Ber,Rev} and
references therein) whereby the energies take
on discrete Matsubara values, $q_0= i n \pi T $, $n$ being
an  even (odd) integer
for bosons (fermions). The usual zero-temperature ultraviolet (UV)
singularities are regularised by dimensional continuation
\cite{DRU}. For simplicity we restrict ourselves
to the Feynman gauge but renormalization via minimal subtraction
\cite{tH} ensures that the coupling
constant is gauge-fixing independent \cite{CW}, and hence
 so will then be our
final answer for the pressure \cite{B}.

At intermediate stages of the calculation we will encounter
various types of infrared (IR) singularities. The first kind is due
 to many-body effects and gives
rise to power-like singularities in some diagrams.
As is well known, these
are removed when the static electric propagator is dressed to
take into
account the screening of electric fields in a plasma
 \cite{GB,AP,Rev}. As a
result of this resummation the
expansion for the pressure is  in powers of $\sqrt{e^2}$
rather than $e^2$,
the famous $e^3$ ``plasmon'' of Gell-Mann and Brueckner \cite{GB}
 illustrating this
at the lowest order.  In this paper we will obtain
also the  order $e^5$
plasmon contribution.

We  remind the reader that the identification of bubble (i.e. no
external legs) diagrams which require the use of screened propagators
so as to produce a consistent perturbative expansion
is most easily done
in the imaginary-time formalism where only the zero mode of the
photon propagator lacks an infrared cutoff of order $T$.
More generally one can perform
a consistent resummation in imaginary-time for any Greens function
whose external legs are bosonic and static (at zero energy).
For Greens
functions which have external fermionic lines, or nonstatic external
bosonic lines, one must first
analytically continue \cite{BM,Rev} to real-time to
obtain the physical
Greens function before the resummation
can be discussed. In the latter case it is in
general necessary to use
nontrivial propagators and vertices to restore the
perturbative expansion,
as discussed by Braaten and Pisarski \cite{BP}.

Though individual bubble diagrams in imaginary time may be
IR finite (after
using dressed propagators if need be), once the frequency sums are
performed the diagram in general splits into several
pieces (integrals)
each of which  individually may contain mass-shell and/or
 collinear singularities.
Since it is convenient to evaluate the different
integrals separately, a
regularisation has to be used for these ``spurious''
IR divergences. We
will use dimensional regularisation (DR) for this also
\cite{DRI,FM,B}.
Actually, in this paper we will
make no attempt to distinguish between IR and UV
divergences at each stage but will only verify that sums of diagrams
which {\it a priori} should be finite, are indeed so. Such finite sums
form gauge-invariant subsets, contributing to the pressure amounts
which are conveniently labelled by different powers of $N$,
the number
of fermion flavours.

We mention here also two other difficulties, related to IR divegences,
in evaluating some integrals. As in the case of the 3-loop pressure
in $\phi^4$ theory \cite{FST}, we  find some integrals containing
singularities along the path of integration which are intrepreted
in the principal value sense.
In some cases we will transform, by a change of variables in the
integral split at the point of singularity, these principal value
singularities into infrared singularities and
then evaluate them by dimensional regularisation. Also some
difficult 3-loop
integrals which are UV finite because of statistical distribution
factors,
but contain a collinear singularity, are in this paper
handled by using
Sudakov variables \cite{suda} in dimensional regularisation
(see \cite{Baggio2}).

 Our notation is as follows : the  wave-vector,
$Q_\mu = (q_0,\vec{q})$ is
contracted with a Minkowski metric, $Q^2=q_0^2 - \vec{q}^2$,
and the measure
of loop integrals is denoted compactly by

\begin{eqnarray}
\int [dQ]\equiv T \sum_{q_0,even}\int {d^{D-1}q\over
(2 \pi)^{D-1}} \, , \nonumber
\end{eqnarray}

\begin{eqnarray}
\int\{dQ\}\equiv T\sum_{q_0,odd}\int {d^{D-1}q\over
(2 \pi)^{D-1}} \, . \nonumber
\end{eqnarray}
The fermions are kept as four-component objects,
$Tr(\gamma_{\mu}\gamma_{\nu})=4 g_{\mu\nu}$, and the gauge
propagator is  $ D_{\mu \nu}(K) =g_{\mu\nu}/K^2$.

In the next section
we review the calculation of the pressure in massless QED to third
($e^3$) order. In Sect.3 the diagrams and integrals which
contribute at
fourth order are discussed. Some frequency-sums are
evaluated in Sect.4
 while in Sect.5 we explain our use of Sudakov variables
to evaluate two
difficult integrals. The final pressure to order $e^4$ is
summarised in
Sect.6  and the renormalisation group briefly discussed. In Sect.7 we
rederive the $e^5$ contribution to the pressure. The results
of this paper
are summarised and discussed in Sect.8 while the appendices
contain some
useful identities and technical derivations.

\setcounter{equation}{0}
\section{Lower Orders}

Before discussing the 3-loop calculation, let us briefly
review the lower order
results for the pressure of QED with $N$ massless Dirac fermions. The
ideal gas  pressure $P_0$ due to electrons, positrons and photons is
given by,
\begin{eqnarray}
 P_{0}&=& {T \over V}   \ln \left\{ \left[
Det_{+} (\partial ^{2}  g_{\mu \nu}) \right]^{-{1 \over 2}} \
 Det_{+}(\partial ^2) \ Det_{-}(i \not\!\! \partial)  \right\}
\label{det} \\
&& \nonumber \\
&=&  \left[ (D-2) + {7 \over 8} (4 N) \right] { T^4 \zeta(4)
\over \pi^2 }
 \nonumber \\
&& \nonumber \\
&=& {\pi^2\over 45} \ T^4 \ \left( 1 + {7\over 4}N\right) \, .
\nonumber
\end{eqnarray}
In (\ref{det})  the $\pm$ subscripts refer respectively to
periodic and antiperiodic boundary conditions, and the second
 determinant
in  is the ghost contribution which is required for proper
counting of physical degrees of freedom in the ideal gas
pressure \cite{Ber}.
The first correction $P_2$ is given by the two-loop diagram of Fig.1,

\begin{eqnarray}
P_2 &=& - {\mu^{4-D} e^2 N \over 2} \int { \{ dK \}
[ dQ ] \ Tr ( \gamma_{\mu} (\not\!\! K-\not\!\!P) \gamma_{\nu}
\not\!\!K  D^{\nu \mu}(P) )
\over K^2 (K-P)^2 } \,  \nonumber \\
&& \nonumber \\
&=& (D-2) \ e^2 \ N \ f_{1} (2 b_1 - f_1 ) \ \mu^{4-D} \ T^{2D-4}
 \label{giraffe} \\
&& \nonumber \\
&=& -{5 e^2 T^4 N\over 288} \, . \nonumber
\label{2c}
\end{eqnarray}
The last line above follows from the previous one as $D \rightarrow 4$.
In the rest of this paper we will often take this limit, where
consistently possible, without comment.  In (\ref{giraffe}), $e$ is the
renormalised
coupling, $\mu$ the mass-scale
of dimensional regularisation and the integrals $b_1$ and $f_1$ are
defined through
\begin{eqnarray}
b_n &\equiv& \int  {[d\,Q]\over (Q^2)^n} \, , \nonumber \\
&& \nonumber \\
f_n &\equiv&  \int { \{d\,Q\} \over (Q^2)^n} \, .  \nonumber
\end{eqnarray}
We have scaled out the temperature in the above integrals
so that $T=1$ there.
The simplest way to evaluate $b_n$ is to first perform
the momentum integrals
and then the frequency sums \cite{BFT}. This gives
\begin{eqnarray}
b_n &=&  {2 \ (-1)^n \ \pi^{{D-1 \over 2}} \over
(2 \pi)^{ 2n} \ \Gamma(n) }
\, \zeta(2n +1 -D) \ \Gamma\left({2n+1-D \over 2}\right)
\, . \nonumber
\end{eqnarray}
Then  $f_n$ is easily obtained by using  a scaling
argument \cite{AE} : Consider the sum $f_n + b_n$ and
 scale the momenta
in the integrals by a factor of $2$. Thus $f_n +b_n =
(2^{2n+1-D}) b_n$,
and so
\begin{eqnarray}
f_n &=& (2^{2n+1-D}-1) \ b_n  \, . \nonumber
\end{eqnarray}

The next contribution to the pressure is
of order $(e^2)^{3/2}$ and reflects
Debye screening \cite{GB,AP,FM,B,Rev}.
To lowest order the static electric polarization tensor
$\Pi_{00}(0,q \rightarrow 0)= m^2 + O(q^2)$
 with $m = e T N^{1/2} /\sqrt{3}$.
Insertions of $\Pi_{00}(0,0)$ along the photon line in Fig.1
create IR divergences which sum to

\begin{eqnarray}
P_3 &=& {T\over 2} \int {d^3 q\over (2 \pi)^3}  \sum_{p=2}^{\infty}
{(-1)^p\over p}
\left( {m^2\over q^2}\right)^p \label{3a}  \\
&& \nonumber \\
\,\,\,\,\,\, &=& - {T\over 2}\int {d^3q\over (2 \pi)^3}
\left( \ln(1+{m^2\over q^2}) -{m^2\over q^2} \right)
\label{3b} \\
&& \nonumber \\
&=& {T\over 2} \  \Gamma \left( {1-D \over 2}\right) {\left(m^2 \over
4 \pi\right)}^{(D-1)/2} \,  \nonumber \\
\label{2d}
&& \nonumber \\
 &=& {e^3 T^4 \over 12 \pi}\left(N \over 3 \right)^{3/2}
\, . \label{3end}
\end{eqnarray}
Though (\ref{3b}) is finite,  we have evaluated it \cite{CoPa}
using dimensional  continuation $3 \rightarrow D-1$.
Then the second term in (\ref{3b})
vanishes and the first term gives the result (\ref{3end})
as $D \rightarrow 4$.
The point of this excursion is to verify, in this example,
that scaleless
integrals like those above may indeed be consistently dropped in DR.
This fact will be useful later.

It is of interest to note that
one may also
calculate the plasmon contribution in real-time but then
the analysis is
much more intricate. In a real-time analysis, the transverse photons
are relevant at intermediate stages but their net contribution
to the $e^3$ plasmon term vanishes \cite{Plas,PlasP} and one recovers
the result of the imaginary-time analysis where
it is clear from the outset (by power counting) that only the
longitudinal photons contribute.
In the language of Braaten and Pisarski \cite{BP}, for the
calculation of the pressure of QED in imaginary time
(recall the discussion
in Sect. 1), the only soft
line is the static photon line. Thus resummation
(which gives the $e^3, e^5$ etc. terms) involves
dressing this soft line  with the relevant static ``hard thermal loop''
which is just the electric mass $\Pi_{00}(0,0)$ \cite{BP}.
For more discussion comparing the real and imaginary-time
approaches to the
calculation of the pressure see \cite{Plas,PlasP}, in particular
the conclusion of the second reference of \cite{Plas}.

\setcounter{equation}{0}
\section{Three Loop Diagrammatics}

The order $e^4 N$
diagrams are shown in Fig.~2 and 3.
For massless fermions, the Ward identity $Z_1=Z_2$
implies
the mutual cancellation of the counterterm diagrams
(Fig.3), so  the sum
$G_1\,+\,G_2$ (Fig. 2) is UV finite.
After performing the spinor traces and  some algebra,

\begin{equation}
G_1=\left(-{e^4 N \over 2}\right) T^{3 D-8} \ \mu^{8-2 D}\  2 \
(D-2)^2 \ \left( f_2(f_1-b_1)^2 + I_2 - 2 I_5\right),
\label{G1}
\end{equation}

\begin{equation}
G_2=\left(-{e^4 N \over 4}\right) T^{3 D-8} \ \mu^{8-2 D} \ 2 \
(2-D) \ \left((8 b_1-16 f_1) I_1 +2 (D-4) I_2 + 8(D-3) I_3
+ 4 I_4\right).
\end{equation}
As before, we have scaled all the
momenta by $1/T$ so that the integrals are dimensionless
(i.e. $T=1$ there). The integrals $b_n$ and $f_n$ were defined in the
last section, while the rest are

\begin{eqnarray}
I_1 &=& \int { \{d\,K d\,R\}\over K^2 R^2 (K+R)^2} \, , \nonumber \\
&& \nonumber \\
I_2 &=& \int {\{d\,K\} [d\,Q d\,P]\over K^2 Q^2 P^2 (K+Q+P)^2}
 \, , \nonumber \\
&& \nonumber \\
I_3 &=& \int {\{d\,K d\,R \ d\,S\}\over K^2 R^2 S^2 (K+R+S)^2}
 \, , \nonumber \\
&& \nonumber \\
I_4 &=& \int {\{d\,K\} [d\,P d\,Q] \ P^2\over K^2 Q^2 (K+Q)^2
 (K+P)^2 (K+P+Q)^2}
\, , \nonumber \\
&& \nonumber \\
I_5 &=& \int {\{d\,K\}[d\,Q\,d\,P] \
     (P\cdot Q)\over K^2 P^2 Q^2 (K+Q)^2 (K+P)^2} \, .\nonumber
\end{eqnarray}

Some simplification is possible. Firstly, within DR, $I_1 =0$
as shown by
Arnold and Espinoza \cite{AE} using scaling arguments. We will
deduce the same
result by direct evaluation in the next section. Using scaling
arguments as
in \cite{AE} one may also show

\begin{equation}
I_2={1\over 6}\left( 2^{11-3 D}-1\right)H_1 -{1\over 6} I_3
\end{equation}
where
\begin{eqnarray}
H_1 =\int {[dQ\,d\,P\,d\,K]\over K^2 Q^2 P^2 (K+Q+P)^2} \nonumber
\end{eqnarray}
is the integral analysed by Frenkel, Saa and Taylor \cite{FST}
 in their
3-loop pressure calculation in hot $\phi^4$ theory. Next,
some algebraic
rewriting gives for $I_4$,

\begin{eqnarray}
I_4 &=& \int {\{d\,K\ d\,R d\,S\} \ (K+S)^2 \over K^2 R^2 S^2
(K+R)^2 (K+R+S)^2}
\nonumber \\
&& \nonumber \\
&=& 2 \ f_1 \ I_1 - {1 \over 2} I_3 \nonumber \\
&& \nonumber \\
&=& - {1 \over 2} I_3 \nonumber \, .
\end{eqnarray}
The last line is  valid because $f_1$ is finite near $D=4$ but
 $I_1 =0$
as mentioned above. Relabelling $I_3 \rightarrow H_2$ and
$I_5 \rightarrow H_3$, we obtain

\begin{eqnarray}
{G_1 + G_2 \over e^4 \ N \ T^{3 D-8} \ \mu^{8-2 D}}& =& { (D-2)
\over 6} \left(\ 2 \ (1-2^{(11-3D)}) \ H_1 \ + \ (20-3D) \
H_2 \ \right) \nonumber \\
&& \ + \ (D-2)^2 \ ( \ 2 \ H_3 \ - \ f_2 \ (f_1-b_1)^2 ) \, .
\label{p1}
\end{eqnarray}

Consider now the order $e^4 N^2$ diagrams.
The photon wave-function
renormalization required for  $G_3$ (Fig.~4a) is provided  by
$X_1$ (Fig. 4b).
Diagram $G_3$ also has an IR singularity which
contributes to the $e^3$ plasmon as the
 first term of the series in eq.(\ref{3a}).
In principle  this term  should be
subtracted from $G_3$ to avoid overcounting but since,
as discussed earlier, it vanishes in DR,
double-counting is automatically avoided.
We have

\begin{eqnarray}
G_3 &=& {e^4 \ N^2 \over 4} \ T^{3 D-8} \ \mu^{8-2 D} \ 16 \ \left(
(D-4) \ b_2 \ f_1^2 + {(D-4)\over 4} \ H_2 + 4 \ H_4\right) \, ,
\label{8a}
\end{eqnarray}
and
\begin{eqnarray}
X_1=-(Z_3 -1) \ e^2 \ N \ (D-2) \ T^{3 D-8} \ \mu^{8-2 D} \ f_1 \
(2 b_1 - f_1) \
\left(T\over \mu\right)^{4-D} \; , \label{x1}
\end{eqnarray}
where
\begin{eqnarray}
Z_3-1={e^2 \ N \over 6 \pi^2 (D-4)} + O(e^4). \nonumber
\end{eqnarray}
In (\ref{8a}) we have dropped a vanishing contribution proportional
 to $I_1$.
The new integral in
(\ref{8a}) is
\begin{eqnarray}
H_4=\int {[d\,Q]\{d\,K\,d\,R\} \ (K\cdot R)^2\over Q^4 K^2 R^2
(Q+K)^2 (Q+R)^2} \ . \nonumber
\end{eqnarray}

In summary, our task for the calculation of the order $e^4$
 contribution
to the presuure as given by eqns.(\ref{p1}, \ref{8a}, \ref{x1})
has been reduced to the
evaluation of the following four integrals :

\begin{eqnarray}
H_1 &=&\int {[dQ\,d\,P\,d\,K]\over K^2 Q^2 P^2 (K+Q+P)^2}
\, , \label{H1} \\
&& \nonumber \\
H_2 &=&\int {\{d\,K d\,R d\,S\}\over K^2 R^2 S^2 (K+R+S)^2}
\, ,\label{H2} \\
&& \nonumber \\
H_3 &=& \int {\{d\,K\}[d\,Q\,d\,P] \ (P\cdot Q)\over K^2 P^2 Q^2
(K+Q)^2 (K+P)^2} \, , \label{H3} \\
&& \nonumber \\
H_4 &=& \int {[d\,Q]\{d\,K\,d\,R\} \ (K\cdot R)^2\over Q^4 K^2 R^2
(Q+K)^2 (Q+R)^2} \ . \label{H4}
\end{eqnarray}
In the next section we will do the frequency sum in $H_1$
and re-obtain
the expression of \cite{FST}. The analysis of $H_2$ is
 completely analogous
to $H_1$. The new integrals we have to analyse are $H_3$ and $H_4$.
These are the main concern of  the next two sections.

\setcounter{equation}{0}
\section{The Frequency Sums}
There are several variations in the literature \cite{Rev,FST}
for performing the frequency sums. Here we sketch one way
(cf. \cite{AB}).\\

{\bf $\bullet I_1$ \\}
Consider the two-loop integral
\begin{eqnarray}
I_1=\int \{d\,K\,d\,Q\}{1\over K^2\, Q^2\, (K+Q)^2}. \nonumber
\end{eqnarray}
The first step is standard and involves writing the $k_0$ sum
as a contour integral,

\begin{eqnarray}
\int \{d\,K\}{1\over K^2 (K+Q)^2}=\int {d^{D-1}\,K\over
(2 \pi)^{D-1}}
\times {1\over 2 \pi i}\oint {-N_{k_0}\over K^2 (K+Q)^2}
\label{ai}
\end{eqnarray}
where
\begin{eqnarray}
N_{k_0}={1\over e^{k_0}+1} \, , \nonumber
\end{eqnarray}
 and the anticlockwise contour
circles the poles of $N_{k_0}$ only. Next the contour is deformed
to enclose only the poles of the propagator. Upon using the
identity
\begin{eqnarray}
{1\over 2 x }\left(f(x) - f(-x)\right)=\int_{-\infty}^{\infty}
d\,x_0 \,\delta(x_0^2-x^2)\epsilon (x_0)f(x_0) \, ,
\label{am}
\end{eqnarray}
and noting that $q_0$ is an odd multiple of $i$,
the result can be written compactly as
\begin{eqnarray}
I_1= \ \int {d^D\ K\over (2 \pi)^{D-1}} \ (N_{k_0} - n_{k_0})  \
\epsilon(k_0) \, \delta(K^2) \int \{d\,Q \} \ {1\over K^2 (K+Q)^2} \; ,
 \nonumber
\end{eqnarray}
where
\begin{eqnarray}
n_{k_0}={1\over e^{k_0}-1} \, . \nonumber
\end{eqnarray}
The above procedure is repeated on the $q_0$ sum but note that
$k_0$ is now real and continuous,
\begin{eqnarray}
I_1= \ \int {d^D\ K d^D \ Q \over (2 \pi)^{2D-2}} \ {\epsilon(k_0) \
\delta(K^2) \ \epsilon(q_0) \ \delta(Q^2) \over (K +Q)^2} \
(N_{k_0} - n_{k_0})  \ (N_{q_0} - N_{-k_0 -q_0}) \nonumber \, .
\end{eqnarray}
Using the identities in the appendix and symmetrizing, this becomes
\begin{eqnarray}
I_1= \ \int {d^D\ K d^D \ Q \over (2 \pi)^{2D-2}} \ {\epsilon(k_0) \
\delta(K^2) \ \epsilon(q_0) \ \delta(Q^2) \over (K +Q)^2} \
(N_{k_0} N_{q_0} - n_{k_0} N_{q_0} - n_{-k_0} N_{-q_0} -1/2) \nonumber
\, .
\end{eqnarray}
Now use the identities in the appendix again to write the
expression above in
terms of $n_k$ and $N_q$ where $k= |k_0|$ and $q=|q_0|$. Then terms
which are independent of statistical factors contribute zero in DR,
terms with only one statistical factor also vanish by Lorentz
covariance and
DR, while terms with two statistical factors vanish in DR
after the energy
integrals are done. Thus $I_1=0$ in agreement with \cite{AE}.\\

{\bf $\bullet H_1$ and $H_2$ \\}
When applied to $H_1$, the frequency sum algorithm discussed
above yields
\begin{eqnarray}
H_1&=&\int {[dK\,\,dQ\,\,dP]\over {K^2 Q^2 P^2 (K \ + Q \ +P)^2}}
\nonumber \\
&& \nonumber \\
&=&\,\,-\int{d^DK\,\,d^DQ\,\,d^DP\over(2 \pi)^{3(D-1)}}
{\epsilon(k_0)\delta(K^2)\,\,\epsilon(q_0)\,\,\delta(Q^2)\,
\,\epsilon(p_0)\,\delta(P^2)\over(K+P+Q)^2} \times
 (4 n_{k_0}n_{q_0}n_{p_0}+2 n_{k_0}n_{p_0})\nonumber \\
&& \nonumber \\
&=&\,\,\,\, -\int {d^DK\,\,d^DQ \,\,d^DP\over (2 \pi)^{3(D-1)}}
{\delta(K^2)\,\,\delta(Q^2)\,\,\delta(P^2)
(4 n_k n_q n_p +6 n_k n_p) \over (K+P+Q)^2} \label{h1}
\end{eqnarray}
where $k = |k_0|$ and we have skipped a few steps which
 make use of the
identities in the appendix. We have also dropped
terms which vanish in DR. To write
(\ref{h1}) in the form of Ref. \cite{FST}, use
\begin{eqnarray}
\int d^{D-1}q\int_{-\infty}^{\infty}d\,k_0\,\,d\,q_0\,\,d\,p_0
{\delta(K^2)\delta(P^2)\delta(Q^2)\over (K+Q+P)^2}
& \equiv &{1\over 2 \pi i}\int d^{D-1}q \int_{-\infty}^{\infty}
d\,k_0\,d\,q_0\,d\,p_0\nonumber \\
&&\,\,\,\,\times{\delta(K^2)\delta(P^2)\over (Q^2-i\,0^+)
((K+Q+P)^2-i\,0^+)} \, , \nonumber  \\
\end{eqnarray}
to get
\begin{eqnarray}
H_1 &=&{-6\over 2 \pi i}\int {d^D K\,d^DQ \,d^DP\over
(2 \pi)^{3(D-1)}}
{\delta(K^2)\delta(P^2)n_k\,n_p\over (Q^2 -i0^{+})((K+Q+P)^2 -i0^+)}
\nonumber \\
&& \nonumber \\
&& \,\,\,\,\,\,\,\,
-4\int {d^DK\,\,d^DQ\,\,d^DP\over (2\pi)^{3(D-1)}}
{\delta(K^2)\delta(Q^2)
\delta(P^2) \, n_k\,\,n_p\,\,n_q \over (K+Q+P)^2} \, . \label{h1t}
\end{eqnarray}
This is the expression of \cite{FST}. Keeping only the
nonvanishing terms
as $D \rightarrow 4$ gives \cite{FST}
\begin{eqnarray}
H_1 &=& { 1 \over 3( 2^6  \pi^2) } \left( 5.6658 - {1 \over (D-4)}
\right) \, .
\label{hh1}
\end{eqnarray}

We leave it as an exercise for the interested reader to show
that $H_2$
(\ref{H2})
is obtained from (\ref{h1t}) by replacing $n_i$ with $-N_i$,
 and then
\begin{eqnarray}
H_2 &=& { 1 \over 3( 2^8  \pi^2) } \left( 2.6045 - {1 \over (D-4)}
\right) \, .
\label{hh2} \\
&& \nonumber
\end{eqnarray}

{\bf $ \bullet H_3$\\}
For $H_3$ we obtain
\begin{equation}
H_3= \int {d^DQ\,\,d^DK\,\,d^DP\over (2\pi)^{3(D-1)}}
\epsilon({q_0})\,\,\delta(Q^2)
\epsilon({k_0})\,\,\delta(K^2)
\epsilon(p_0)\,\,\delta(P^2)\,\,\{{\cal R}\},
\label{I5}
\end{equation}
where

\begin{eqnarray}
\{{\cal R}\}&\equiv & {T_K\over (K+Q)^2 (K+P)^2}
+ {\left(T_K|K\rightarrow K-Q\right)\over (K-Q)^2(K-Q+P)^2} +
{\left(T_K|K\rightarrow K-P\right)\over (K-P)^2 (K-P+Q)^2}, \nonumber
\end{eqnarray}
\begin{eqnarray}
T_K &\equiv& N_{k_0} \left( n_{p_0}[n_{q_0}P\cdot Q
-N_{-q_0}P\cdot (Q+K)]\right.
\nonumber \\
&&\,\,\,\,\,\,\left.
-N_{-p_0}[n_{q_0}Q\cdot(P+K)- N_{-q_0}(Q+K)\cdot (P+K)] \right),
\nonumber
\end{eqnarray}
and  $\left(T_K| K \rightarrow K-Q\right)$ means {\em replace}
$K_\mu$ by
$K_\mu- Q_\mu$ in the expression for $T_K$. After the usual
 simplification,
\begin{equation}
H_3=J_1+K_1+L_1. \label{h3}
\end{equation}

The piece $L_1$ contains integrals that can be performed
exactly using relations (\ref{r1})-(\ref{ar}) of  Appendix A.

\begin{eqnarray}
L_1&=&{\omega(D)\omega^2(D-1)\over (2 \pi)^{3(D-1)}} \ 2^{(D-5)} \
B(1/2, \ D/2-1)\,\,B(D/2,\ D/2-2) \nonumber \\
&& \nonumber \\
&&\,\,\,\,\,\, \times {\cal M}_2 (D-5) \ \left[{\cal M}_1
(D-3)+ {\cal M}_2
(D-3)\right]^2 \nonumber \\
&& \nonumber \\
&&\,\,\,\,\, + {\omega (D)\omega (D-1)\over (2 \pi)^{3(D-1)}}\,\,
B(D/2-1, \ D-3) \ 2^{D-6}\,\, {\cal N}_1, \nonumber
\end{eqnarray}
where
\begin{eqnarray}
{\cal N}_1&=& 3 {\cal M}_1 \left({3 D-10\over 2}\right)
\ {\cal M}_2 \left({3 D-10 \over 2}\right) \sum_{\sigma,\gamma=\pm 1}
(\sigma  \gamma) C_0(\sigma,\gamma) \nonumber \\
&& \nonumber \\
&&\,\,\,\,\,\,  +2 \left[{\cal M}_1 \left({3 D-10\over 2}\right)+
{\cal M}_2
\left({3 D-10\over 2}\right) \right]^2
\sum_{\sigma,\gamma=\pm 1}C_1(\sigma,\gamma). \nonumber
\end{eqnarray}
The functions $ \omega$, ${\cal M}_i$ and $C_i$ are defined in the
appendix and $B(a,b)$ is the usual beta-function.
Expanding about $D=4$,
\begin{eqnarray}
L_1 &=& {-1 \over 3  (2^9  \pi^4) }
\left({2\pi^2 + 7 \pi^2 \ln \,2 + 6\pi^2 \zeta'(0)+ 18 \zeta'(2)
\over (D-4)} + 35.478 \right). \nonumber \\
\end{eqnarray}

In (\ref{h3}) $K_1$ is an integral similar to that appearing in
$H_1$ and $H_2$,
\begin{eqnarray}
K_1 &=&-\int {d^4K\,\,d^4Q\,\,d^4P\over (2\pi)^9}
{\delta(P^2)\delta(Q^2)\delta(K^2)N_p\,N_q\,n_k \over (K+P+Q)^2}
\nonumber \\
&& \nonumber \\
&=&{1.1439\over 2 (2 \pi)^6}.
\end{eqnarray}

Finally, $J_1$ is defined by
\begin{eqnarray}
J_1&=&\int {d^DK\,\,d^DQ \,\,d^DP \over (2 \pi)^{3(D-1)}}
\delta_+(K^2)\delta_+(Q^2)\delta_+(P^2)N_{k_0}n_{q_0}(N_{p_0}+n_{p_0})
\sum_{\sigma,\gamma=\pm 1}(-\sigma)\,\,{\cal S}_1 (\sigma,\gamma),
\nonumber \\ \label{jj1}
\end{eqnarray}
with
\begin{equation}
{\cal S}_1 (\sigma,\gamma)={P\cdot Q\over K\cdot Q}{1\over K\cdot Q +
P\cdot(\sigma K +\gamma Q)} \, ,
\end{equation}
and $\delta_{+}(Q^2) = \delta(Q^2) \theta(q^0)$. The evaluation of
$J_1$ is  described in Sect.5.

Collecting the pieces,
\begin{eqnarray}
H_3&=&J_1+ K_1 + L_1\nonumber \\
&& \nonumber \\
&=&{1\over 2^7 \pi^4}\left( 1.095 -{0.4112335165\over (D-4)} \right).
\label{hh3} \\
&& \nonumber
\end{eqnarray}

{\bf $ \bullet H_4$\\}
The only extra point here is the doubled propagator $1/(Q^2)^2$.
Thus the contribution to the contour integrals coming from
this propagator
must be extracted correctly using the appropriate formula
of complex analysis.
Other than this technical point, the rest of the derivation
proceeds as
discussed for $H_3$ but the expressions are lenghtier.
So we will only
state the results.

We obtain
\begin{equation}
H_4=J_2 + K_2 + L_2,
\end{equation}
where $J_2$ is an integral similar to $J_1$ and is discussed in the
next section.
\begin{eqnarray}
K_2 &=&{1\over 2}\int {d^4 K\, d^4 Q \,d^4 P\over (2 \pi)^9}
{\delta(K^2)\delta(Q^2)\delta(P^2) \ N_k \ N_q \ N_p
\over (K+P+Q)^2}\nonumber \\
&& \nonumber \\
&=&{-0.4417\over 4 (2\pi)^6},
\end{eqnarray}
and
\begin{eqnarray}
L_2 &=& {1 \over 3^3 \ 2^{10} \ \pi^4 }
\left({17\pi^2 +12 \pi^2 \ln \,2+24\pi^2 \zeta'(0)+72 \zeta'(2)
\over (D-4)} - 0.5098 \right). \nonumber \\
\end{eqnarray}

Finally
\begin{eqnarray}
H_4 &=& J_2 +K_2 + L_2\nonumber \\
&& \nonumber \\
&=&{1\over 2^7 \pi^6}\left(1.407 -{2.254840072\over (D-4)}\right).
\label{hh4}
\end{eqnarray}
\\

\setcounter{equation}{0}
\section{Sudakov Variables}
In this section we describe the evaluation of the integrals
 $J_1$ and $J_2$
appearing in the last section.

{\bf $ \bullet J_1$\\}
The integral $J_1$ was
defined as (\ref{jj1}),
\begin{eqnarray}
J_1&=& \sum_{\sigma,\gamma=\pm 1}(-\sigma)
\int {d^DQ \,\,d^DP \over (2 \pi)^{3(D-1)}} \
\delta_+(Q^2) \ \delta_+(P^2) \ n_{q_0} \ (N_{p_0}+n_{p_0})
\nonumber \\ \label{jj11}
&& \;\;\;\;\;\;\;\;\;\;\;\;\;\;\;\; \times
\int d^DK \ \delta_+(K^2) \ {P\cdot Q\over K\cdot Q}
{N_{k_0}\over K\cdot Q +P\cdot(\sigma K +\gamma Q)} \, .
\end{eqnarray}
This integral is quite complicated and
we have not succeeded in
evaluating it in closed form. However for our purposes we only require
the order $(D-4)^{-1}$ and $O (D-4)^{0}$ terms from (\ref{jj11}) in the
limit $D \rightarrow 4$. Although (\ref{jj11}) is UV finite, a
$1/(D-4)$ pole is expected because of the collinear singularity
as $K.Q \rightarrow 0$.

Let us concentrate on the $K$-subintegral appearing in (\ref{jj11}),
\begin{equation}
\int d^DK \ \delta_+(K^2) \ {P\cdot Q\over K\cdot Q}
{N_{k_0}\over K\cdot Q +P\cdot(\sigma K +\gamma Q)} \, .
\label{egg}
\end{equation}
For the rest of this subsection we will mostly discuss this subintegral with
the implicit understanding that it occurs inside (\ref{jj11}) so that the
constraints $P^2=Q^2=0$ and $p_0=p, \  q_0 =q$ hold. It is sufficient
to extract only the
$O (D-4)^{-1}$ and $O (D-4)^{0}$ pieces from (\ref{egg}) because the
remaining $(P,Q)$ integrals in ({\ref{jj11}) are UV and IR finite ( so that
we will only need their $O (D-4)^0$ and $O (D-4)^{1}$ pieces to get the
full $O (D-4)^{-1}$ and $O (D-4)^{0}$  terms for $J_1$ when our result for
(\ref{egg}) is substituted back into (\ref{jj11}) ).

Since the integral (\ref{egg}) involves three independent scalar products
in $D$ dimensions, it poses a formidable problem. We begin our task by
decomposing the loop momentum $K$ in a Sudakov base \cite{suda} constructed
on $P$ and $Q$ in such a way that the $D$ dimensional angular variables
will disappear from the denominator in (\ref{egg}). Thus we write
\begin{equation}
K \equiv \alpha P + x Q + K_\perp
\label{base}
\end{equation}
with $K_\perp$ denoting a transversal vector
($P \cdot K_\perp=Q \cdot K_\perp=0$)
of  $D-2$ nonzero components. Note that the
 delta and step-functions in (\ref{jj1})
imply $K_{\perp}^2 \le 0$. Since the Jacobian in going from the
$K-$basis to the $(\alpha, x, K_\perp)$ basis is $s \equiv P.Q$, we obtain for
(\ref{egg}) (upon using (\ref{base}) to eliminate $P.K$ and $Q.K$ ),
\begin{eqnarray}
&& \int_{-\infty}^{\infty} d \alpha \int_{-\infty}^{\infty} d x \int
d^{D-2} K_\perp \,\, { \delta(2sx\alpha + K_{\perp}^{2}) \over \alpha} \; \,
{ N_{k_{0}} \ \theta(k_{0}) \over \alpha + (\sigma x + \gamma) } \\ \label{s1}
&& \nonumber \\
&=& \int_{-\infty}^{\infty} d x \int d^{D-2} K_\perp \;
\, { 1 \over (-K_{\perp}^{2})} \; \,
{ N_{k_0} \, \theta(k_{0}) \over {(-K_{\perp}^{2}) \over 2sx} +
(\sigma x + \gamma)}
\label{s2} \, ,
\end{eqnarray}
where now
\begin{equation}
k_{0} = { \left(-K_{\perp}^{2}\right) \over 2sx} p +xq + (K_{\perp})_{0} \ ,
\label{ko}
\end{equation}
which follows from (\ref{base}) and the fact that $p_0=p$ and $q_0=q$.
The original collinear singularity of the $K-$integral now appears as
an endpoint IR singularity as $D \rightarrow 4$ in the
radial part of the $K_\perp$-integral.
This is not surprising since we note from (\ref{base}) that
the collinear singularity of (\ref{egg})
is encountered on the plane spanned by the two massless
momenta $P$ and $Q$.
The region of integration which contributes to this singularity is
therefore identified by the limit $K_\perp \rightarrow 0$.

The combination of Sudakov methods and dimensional regularisation
to study discontinuities in { \it zero} temperature type integrals
is discussed in \cite{Baggio2}. In the present case though we have
obtained a simplified structure for (\ref{egg})
in terms of (\ref{s2}), the dependence of the statistical factor
on $k_0$ is still a complication.
We therefore  isolate the  IR singularity in (\ref{s2})
into a simpler integral by rewriting
\begin{equation}
N_{k_{0}} \ \theta(k_{0}) = [ N_{k_{0}} \ \theta(k_{0}) -
N_{xq} \ \theta(xq)] \, + N_{xq} \ \theta(xq).
\label{split}
\end{equation}
The reason for this rearrangement is because now
the term in square-brackets on the right-hand-side above
gives an UV and IR finite
\footnote{The reader who does not find this too apparent may find
{\it a posteriori} satisfaction by examining the simplified form of
the relevant equation in Appendix B.}
contribution to (\ref{s2}) ( and hence to $J_1$) as $D\rightarrow 4$,
while the simpler last term in (\ref{split}) will give the pole
( and some finite parts). Calling the net contribution of
the square bracket in (\ref{split}) to $J_{1}$ as $J_{1A}$,
we obtain (see Appendix B)
\begin{equation}
J_{1A}={6.101 \over 2^7 \pi^6}.
\label{5.7}
\end{equation}

The last term on the right-hand-side of (\ref{split}) contributes to
(\ref{s2}) the amount
\begin{equation}
\int_{-\infty}^{\infty} d x \int d^{D-2} K_\perp \,
{ 1 \over (-K_{\perp}^{2})} \,
{ N_{xq} \, \theta(xq)  \over {(-K_{\perp}^{2}) \over 2sx} +
(\sigma x + \gamma)}
\label{s3} \, .
\end{equation}
Since $K_\perp$ is a space-like vector with
$D-2$ independent components we may write
$K_\perp = \sum_{i=1}^{D-2} \sigma_{i} \ e_{i}$,
where the $e_{i}$ are orthonormal space-like vectors :
$e_{i} \cdot e_{j} = - \delta_{ij}$. Then (\ref{s3})
becomes

\begin{eqnarray}
&& \int_{-\infty}^{\infty} dx \int_{-\infty}^{\infty}
 {d\sigma_{1}.....d\sigma_{D-2}\over (\sigma_{1}^2 +......+ \sigma_{D-2}^2)}
{ N_{xq} \ \theta(xq) \over (\sum \sigma_{i}^2 /2sx) + (\sigma x + \gamma)}
\label{sug} \\
&& \nonumber \\
&=& \omega(D-1) \int_{-\infty}^{\infty} d x \int_{0}^{\infty} dr \ r^{D-5} \,
{ N_{xq} \, \theta(xq)  \over {r^2 \over 2sx} + (\sigma x + \gamma)}
\label{suc} \\
&& \nonumber \\
&=& { \omega(D-1) \over 2} \int_{0}^{\infty} d x \int_{0}^{\infty}
 dy \ y^{D/2-3} \, { N_{xq}  \over {y \over 2sx} + (\sigma x + \gamma)} \\
&& \nonumber \\
&=& {\omega(D-1) \over 2} \ (2 \,s)^{D/2-2} \, \int_{0}^{\infty}
dx {x^{D/2-2}\over e^{x q}+1}\int_{0}^{\infty}dz{z^{D/2-3}\over
z +(\sigma x+\gamma)} \; .
\label{y2part}
\end{eqnarray}
In going from (\ref{sug}) to {\ref{suc}) we switched from the
Cartesian coordinates $\sigma_{i}$ to spherical coordinates and the
function $\omega(D-1)$ as defined by (\ref{om}) is the result
of doing the angular integrals while $\int dr$ is the remaining radial
integral.
When (\ref{y2part}) is substitued back into $J_1$,
we will need the sum
\begin{equation}
\sum_{\sigma,\gamma = \pm 1 }(-\sigma){1\over z+(\sigma x + \gamma)}
=\sum_{\sigma = \pm 1}(-\sigma)\left({1\over z+\sigma(x+1)}+
{1\over z +\sigma(x-1)}\right).
\label{cstar}
\end{equation}
The change of variables $z=y(x+1)$ for the first term in
(\ref{cstar}) and $z=y|x-1|$for the second term decouples the
$x-z$ integral in (\ref{y2part}) into a product of an $x$-integral and
a simple $y$-integral. The resulting $y$-integrals can be evaluated
explicitly using eqns.(\ref{aa29}-\ref{ar}). The $x$-integrals are still
too complicated to be evaluated in closed form but the
their integrands may be
expanded about $D=4$ directly or after some integration by parts.
In this way we obtain the net contribution, $J_{1B}$,
 of (\ref{y2part}) to $J_1$ as
\begin{equation}
J_{1B}= \int_{0}^{\infty} dq \ q \ n_{q} \left\{
{4 g_{0} \over (D-4)} [\mbox{Box} 1]
 + g_{0} \left( 2 [\mbox{Box} 2] + 6 \ln(q) [ \mbox{Box} 1] \right) +
4 g_{1} [\mbox{Box} 1]
 \right\} + O(D-4) \, ,
\end{equation}
where
\begin{equation}
[\mbox{Box} 1] = -\int_{0}^{\infty} dx {N_{xq} \over x+1} \ +
\int_{0}^{1} dx {N_{xq} -N_{q} \over (1-x) } \ +
\int_{1}^{\infty} dx \ q \ N_{xq}(N_{xq}-1) \ln(x-1) \, \, ,
\end{equation}

\begin{eqnarray}
[\mbox{Box} 2] &=& { -\pi^2 \over 6} N_{q} \
-\int_{0}^{\infty} dx {N_{xq} \over x+1} \ln x(x+1) \ +
\int_{0}^{1} dx {N_{xq} -N_{q} \over (1-x) } \ln x(1-x) \nonumber \\
&& \nonumber \\
&& \;\;\;\;\; + \int_{1}^{\infty} dx \left( {N_{xq} \over x} \ln x(x-1) \ + \
q N_{xq}(N_{xq}-1) \ { [\ln x(x-1)]^2  \over 2} \right) \, \, ,  \nonumber \\
\end{eqnarray}
and
\begin{eqnarray}
g_0 &=& {1 \over 512 \pi^4} \, ,\\
&& \nonumber \\
g_1 &=& {-2 \pi^2 + \pi^2 \ln2 -3 \pi^2 \ln \pi + 18 \zeta^{\prime}(2)
 \over 256 \pi^6} \, .
\end{eqnarray}

Evaluating the integrals above numerically we obtain
\begin{equation}
J_{1B}= {1 \over 2^7 \pi^4} \left( {r_1 \over (D-4)} + 3.3175 \right)
\end{equation}
with
\begin{eqnarray}
r_1&=&\int_{0}^{\infty}dq\,\,q\,n_q [\mbox{Box} 1] \\
&& \nonumber \\
&=&-0.7167667897.... \; .
\end{eqnarray}

Thus finally
\begin{eqnarray}
J_1&=&J_{1A}+J_{1B}\nonumber \\
&& \nonumber \\
&=& {1\over 2^7 \pi^4}\left( 3.936 -{0.7167667897\over (D-4)}
\right). \\
&& \nonumber
\end{eqnarray}

Before going on to $J_2$, let us recapitulate the story of $J_1$.
Our objective was to obtain the order $1/(D-4)$ and $O(D-4)^0$ pieces
of $J_1$ as $D \rightarrow 4$. We first used a technique popular and
effective at zero temperature, the Sudakov decomposition, which exposed
the singularity structure of $J_1$ in a more manageable form. However,
unlike $T=0$ cases, further progress was hampered by the presence of the
statistical factor which in the Sudakov basis obtains a complicated energy
dependence. Our next step was to isolate the singularity of $J_1$ into
a simpler integral by adding and subtracting terms in the original
integral (see (\ref{split})).
The simpler part of $J_1$, which we called $J_{1B}$, had a  statistical
factor which did not depend on the transversal momentum $K_\perp$ of the
Sudakov decomposition and so we were able in this case to proceed further
in extracting the singular and finite pieces of $J_{1B}$ as $D \rightarrow 4$.
The other part of $J_1$, which was created by the shift (\ref{split}) and
which we called $J_{1A}$, was a complicated but finite object
at $D=4$ so we could  evaluate it numerically. \\ \\

{\bf $ \bullet J_2$ \\}
In the evaluation of $H_2$ in Sect. 4 we required,
\begin{equation}
J_2 \equiv J_{2\,1}+J_{2\,2},
\end{equation}
where

\begin{equation}
J_{2\,1}=-\int{d^DK\,d^DQ\,d^DP\over (2 \pi)^{3(D-1)}}
N_k N_qN_p \delta_+(K^2)\delta_+(P^2)\delta_+(Q^2)
\sum_{\sigma\,\gamma = \pm 1}(-\sigma){\cal S}_1(\sigma,\gamma),
\end{equation}
and
\begin{equation}
J_{2\,2}=\int {d^DK\,d^DQ\,d^DP\over (2\pi)^{3(D-1)}}
N_k\, N_q\,N_p\delta_+(K^2)
\delta_+(Q^2)\delta_+(P^2)\sum_{\sigma\,\gamma = \pm 1}(\sigma\gamma)
{\cal S}_2(\sigma,\gamma),
\label{j22}
\end{equation}
and where we have defined

\begin{equation}
{\cal S}_N(\sigma,\gamma)=\left({P\cdot Q\over K\cdot Q}\right)^N
{1\over K\cdot Q + P\cdot (\sigma K + \gamma Q)}.
\end{equation}

The evaluation of $J_{2\,1}$ is similar to that of $J_1$ discussed earlier.
We find
\begin{equation}
J_{2\,1}={0.221 \over 2^7 \pi^6}.
\end{equation}
Notice however that unlike $J_1$ there is no singular $1/(D-4)$
contribution from $J_{2\,1}$.

Consider next the $K$-integral in $J_{2\,2}$ (\ref{j22}) :
\begin{equation}
\int d^D K  \delta_+(K^2)\left({P\cdot Q\over K\cdot Q}\right)^2
{N_{k_0} \over K\cdot Q + P\cdot (\sigma K + \gamma Q)}.
\end{equation}
In the Sudakov basis this becomes
\begin{eqnarray}
 \int_{-\infty}^{\infty} d x \int d^{D-2} K_\perp
\, { 2sx  \over (-K_{\perp}^{2})^2} \; \
{ N_{k_0} \ \theta(k_{0}) \over {(-K_{\perp}^{2}) \over 2sx} +
(\sigma x + \gamma)}
\label{s4} \, ,
\end{eqnarray}
with $k_0$ again given by (\ref{ko}). As $D \rightarrow 4$, the IR
singularity in this $K_\perp$ integral is now more severe than the case of
(\ref{s2}) so that the rearrangement (\ref{split}) by itself is not
sufficient to simplify the integral. However what is eventually
required in (\ref{j22}) is
\begin{eqnarray}
\sum_{\sigma \gamma = \pm 1} ( \sigma \gamma) \, [\mbox{eqn}.(\ref{s4})] \, ,
\label{s5}
\end{eqnarray}
and we see that the IR behaviour of (\ref{s5}) as $K_\perp \rightarrow 0$
is similar to that of (\ref{s2}). Hence the split (\ref{split})
can be used as before to obtain the pole and finite pieces for $J_{22}$.
Since the analysis is very much the same as before (see however some comments
at the end of Appendix B), we simply state the
result :
\begin{equation}
J_{22}={1\over 2^7 \pi^6}\left(1.430 -{0.6420474257\over (D-4)}
\right).
\end{equation}

Thus
\begin{eqnarray}
J_2 &=& J_{21}+J_{22}\nonumber \\
&& \nonumber \\
&=& {1\over 2^7\pi^6}\left(1.651 - {0.6420474257\over (D-4)}\right).
\end{eqnarray}
\\

\setcounter{equation}{0}
\section{Results to Fourth Order }

The pressure up to and including order $e^4$ then follows from
Sect.2 and eqns.(\ref{p1})-(\ref{x1}), (\ref{hh1}),(\ref{hh2}),
(\ref{hh3}),
(\ref{hh4}) :
\begin{eqnarray}
{P \over T^4}&=& {\pi^2\over 45}  \left( 1 + {7\over 4}N\right)
 -{5 e^2 N\over 288} +
 {e^3 \over 12 \pi}\left(N \over 3 \right)^{3/2} \nonumber \\
&& \nonumber \\
&& \;\;\;\;\;\; +  { e^4 N\over \pi^6}(0.4056) -  e^4 N^2
\left( {0.4667 \over \pi^6} + {5\over 6\pi^2 \times 288}\ln
{T\over \mu}\right).
\label{banana}
\end{eqnarray}
The coupling above is an implicit function of the renormalisation
scale $\mu$.
One may choose $\mu=T$ so as to eliminate the
logarithm at this and higher orders. Then the pressure is a
function of
$e(T)$. In principle the value of $e(T)$ may be determined by comparing
the perturbative
calculation of some other observable at super-high temperature $T$
(where the electron mass is negligible) with its
experimentally measured value. Alternatively
one can use the renormalisation group to relate $e(T)$ to the
 coupling
at some other scale $\Lambda$. Perhaps the most instructive
thing to do is to
write (\ref{banana}) in terms of the perturbative
 renormalization-group-invariant coupling, at the
energy scale $T$, given by
\begin{eqnarray}
e^2(T)=e^2\left( 1+ {e^2 N \over 6 \pi^2} \ln {T\over \mu}\right)
+ O(e^6)
\ . \label{tcop}
\end{eqnarray}
Defining $\alpha(T)={e^2(T)/ 4\pi}$ we arrive at
\begin{eqnarray}
P\over T^4 &=& {\pi^2\over 45} \ (1 +{7\over 4}N) \ - \
{5\pi^2\over 72} \ {\alpha(T)N\over \pi} \ + \ {2 \pi^2
\over 9 \sqrt{3} }
\left({\alpha(T) N\over \pi}\right)^ {3/2} \nonumber \\
&& \nonumber \\
&&\,+\left({0.658 \pm 0.006 \over N}
- 0.757 \pm 0.004 \right) \left({\alpha(T) N \over
\pi}\right)^2 + O\left(\alpha(T)^{5/2}\right). \nonumber \\
&& \label{papaya}
\end{eqnarray}

Notice the disappearance of the logarithm. We have also indicated in
(\ref{papaya}) estimates of numerical
uncertainties due to the evaluation of
some integrals by quadratures.

\setcounter{equation}{0}
\section{Fifth Order }
The next correction to the pressure is of order $e^5$ and comes about
by dressing the photon lines of the $3$-loop diagrams. Its
calculation  is completely analagous to that of the
$e^3$ term reviewed in Sect.2 and the reader is encouraged to re-read
the discussion there and in the references quoted. Since
it has already been discussed at length in
\cite{P}, here we only sketch another derivation using the
``ring-summation'' formula.

Consider the static, renormalised, {\it one}-loop
photon polarisation tensor $\Pi_{\mu\nu}(q_0=0,q)$. From gauge
invariance, $Q_{\mu} \Pi^{\mu \nu}(Q) = 0$, one obtains
$\Pi_{i0}(0,q)=0$
 while  explicit calculations yield
$\Pi_{ij}(0,q \rightarrow 0) = O(e^2 q^2)$ and
$\Pi_{00}(0, q \rightarrow 0) = m^2 + O(e^2q^2)$. Thus for
diagrams $G_1$
and $G_2$ (Fig.2), one deduces from
the usual power counting that it is only necessary to dress one of the
photon lines with
the static one-loop electric polarisation tensor to get the
$e^5$ contribution.
The resulting dressed diagarms are of the form of Fig.5 and are
contained in the full ring sum which is summarised by the
formula (see the third reference of \cite{Rev}),
\begin{eqnarray}
P_{ring} = -{1 \over 2} \int [dQ] \ Tr \{ \ln(1-D(Q) \
\hat{\Pi}(Q)) + D(Q) \
\hat{\Pi}(Q) \} \, , \label{ring}
\end{eqnarray}
where the trace, $Tr$, is over Lorentz indices, $\hat{\Pi}$
is the full
self-energy and $D(Q)$ the bare propagator. Define $\hat{F}(q) \equiv
\hat{\Pi}_{00}(0,q)$. Then the restriction of (\ref{ring})
to the static
electric sector (as mentioned earlier this is the only sector which will
give the $e^5$ contribution) gives
\begin{eqnarray}
 {T \over 2} \int {d^{D-1} q  \over (2 \pi)^{D-1} } \
\sum_{n=2}^{\infty} \
{ (-1)^n \over n } \left({ \hat{F}(q) \over q^2} \right)^n \,
. \label{rest}
\end{eqnarray}
Now write $\hat{F}(q) = F^1(q) + F^2(q) + ...$, where the
superscripts refer to the
loop order. Then diagrams like those of Fig.5 are obtained
by truncating
(\ref{rest}) to the appropriate sector,
\begin{eqnarray}
[\mbox{eq.(\ref{rest})}] &\rightarrow &
 {T \over 2} \int {d^{D-1} q \over (2 \pi)^{D-1} } \
\sum_{n=2}^{\infty} \
{ (-1)^n \over n }  { n \ F^2(q) \ (F^1(q))^{n-1}
\over (q^2)^n } \,  \\
&&\nonumber \\
&& \;\;\; \; \; =  {T \over 2} \int {d^{D-1} q  \over (2 \pi)^{D-1} }
  {  F^2(q) \ F^1(q)  \over q^2 \ (q^2 + F^1(q) ) } \, .  \label{pinto}
\end{eqnarray}
By scaling $ \vec{q} = m \vec{x}$ one deduces that for the $e^5$
contribution it is sufficient to set $q=0$ in the $F^i$ of
 eq.(\ref{pinto}).
Thus the order $e^5 N^{3/2}$ contribution to the pressure is given by
\begin{eqnarray}
[\mbox{eq.(\ref{pinto})}] \rightarrow
{T \ \sqrt{F^1(0)} \over 8 \pi} \ F^2(0) &=&
{e^2 T \sqrt{F^1(0)} \over 8 \pi} \ { \partial^2 P_2 \over
\partial \mu^{2}_{e}}
|_{\mu_{e}=0} \,  \label{burp} \\
&& \nonumber \\
&=& {-e^5 T^4 N^{3/2} \over 64 \pi^3 \sqrt{3}}  \label{bburp} \, ,
\end{eqnarray}
where $F^1(0) = m^2 = e^2 T^2 N / 3$ and $P_2$ denotes the pressure
at two-loop
order at chemical potential  $\mu_e$ \cite{Rev}.
In (\ref{burp})
we used the relation $\hat{\Pi}_{00}(0,0) = e^2 \partial^2 P /
\partial \mu^{2}_{e} $ \cite{Rev}. The result (\ref{bburp})
may  also be obtained by direct calculation of the left-hand-side
of (\ref{burp}) \cite{P}.

At order $e^5$ we still have the $e^5 N^{5/2}$
contribution obtained by dressing $G_3$ (Fig. 4a).
Now the ring summation formula (\ref{rest}) has to be
 truncated in the
sector where only iterations of $F^1(q)$ occur, but for one
of them the
subleading momentum dependence is taken while the rest are at zero
momentum :
\begin{eqnarray}
[\mbox{eq.(\ref{rest})}] & \rightarrow & {T \over 2} \int {d^{D-1} q \over
(2 \pi)^{D-1} }
\ \sum_{n=3}^{\infty} \ (-1)^n \
\left({ \partial F^1 \over \partial q^2 }\right)_{q^2 =0}
 \left({ F^1(0) \over q^2}\right)^{n-1}  \\
&&\nonumber \\
&& \; =   {-m^3 T \over 8 \pi}
{ \partial F^1 \over \partial q^2 }|_{q^2 =0} \\
&& \nonumber \\
&& \; ={ -e^5 T^4 N^{5/2} \over  8 \pi \sqrt{27}} \
{\gamma-1 + \ln(4/\pi) \over 12 \pi^2}\ +
\ { e^5 T^4 N^{5/2} \over 8 \pi \sqrt{27}} \ {\ln(T/\mu)
\over 6 \pi^2}  \, .
 \label{chiquita}
\end{eqnarray}
As in the last section, the logarithm eventually disappears
when the pressure
is written in terms of the temperature-dependent
coupling (\ref{tcop}).

 The derivation of (\ref{burp}) may be
extended naturally to re-obtain the identity of \cite{P} linking the
contribution of the pressure at order $2n+3 \ ( n \ge 1) $
from diagrams with one-fermion
loop to the pressure at order $2n$. The relation (\ref{burp})
 and its
higher order extensions were stated in \cite{P} for the case
of massless
fermions at zero chemical potential. Clearly one can relax
these restrictions
since only the dressing of the photon is involved. Thus in general
\begin{eqnarray}
P_{2n+3}^{1F} &=& {e^2 T \sqrt{F^1(0)} \over 8 \pi}
\ { \partial^2 P_{2n}^{1F}
 \over \partial \mu^{2}_{e}}  \,  \label{slurp}
\end{eqnarray}
relates the gauge-invariant pressure at order $2n+3 \ ( n\ge 1)$,
from  diagrams with one fermion loop, of QED with massive
electrons at {\it non}-zero temperature but arbitrary
chemical potential,
to the pressure at order $2n$ (the nonzero $T$ is required so as
to isolate the zero mode to be dressed).
In this general case,
$F^1(0)$ is the lowest order ($e^2$)
electric screening mass at nonzero $T, \mu_e$ and electron mass.

\setcounter{equation}{0}
\section{Conclusion }

With $\alpha(T) = e^2(T) / 4 \pi$, and
defining $g^2 = \alpha(T) N / \pi$,
the pressure of QED with
$N$ massless Dirac fermions at nonzero temperature, $T$,
is given to fifth order by
\begin{eqnarray}
P\over T^4 &=& a_0 + g^2 a_2 + g^3 a_3 + g^4 (a_4 + b_4 / N) +
g^5 (a_5 + b_5 / N) + O(g^6) \, ,
\end{eqnarray}
with
\begin{eqnarray}
a_0 &=& {\pi^2\over 45} \ (1 +{7\over 4}N) \, , \\
a_2 &=& - {5\pi^2\over 72} \, ,\\
a_3 &=&  {2 \pi^2 \over 9 \sqrt{3} } \, , \\
a_4 &=& - 0.757 \pm 0.004 \, , \\
b_4 &=&  0.658 \pm 0.006 \, , \\
a_5 &=&  { \pi^2 [1- \gamma - \ln(4 / \pi)] \over 9 \sqrt{3}} \
= \ 0.11473... \; , \\
b_5 &=&   {- \pi^2 \over 2 \sqrt{3} }= -2.849... \; .
\end{eqnarray}

Real world QED corresponds to $N=1$, but since we have
ignored the electron mass  the results
are applicable only
at extremely high temperatures \cite{KKT}.
Numerically, the fourth and fifth order terms we have
found are  small corrections in the regime where the coupling
itself is small.
However,  since perturbative QED is not
asymptotically free, the effective coupling $\alpha(T)$
increases slowly with temperature  so the results might be
of use for physics of the very early universe  or,
more speculatively, for studies of strongly coupled
QED (some references are in \cite{P}).

It is an amusing fact that the order $e^5$ contribution
to the pressure
of QED is much easier to calculate than the $e^4$ contribution.
 Indeed, as
discussed in the last section, the fifth order calculation
may even be extended
to massive electrons and nonzero chemical potential (but nonzero $T$).
 However we have not
bothered to give the explicit expressions in those cases
because the fourth
order calculation  at nonzero $T$ is itself  only known for
massless electrons
at zero chemical potential \cite{CoPa}.

A 3-loop calculation in QCD will differ from the QED case
in two respects. Firstly there is an
increase in the number of diagrams. This however is not a
problem (except for
tedium) as we feel that our approach
using the frequency-sum algorithm discussed in
Sect.5 and the Sudakov method of Sect.6 is
general enough to handle
any new integrals that might arise.
The second difference is that the static electric polarisation tensor
in QCD behaves as $\Pi_{00}(0,q\rightarrow 0) = M^2 + qT$ and this
gives rise to the $g^4 \ln g$ term \cite{T} from the sum of ring
diagrams. Thus in this case one has to be more careful in using
dimensional regularisation (as in this paper) to extract this term and
also the constant under the logarithm.\\

{\it Note added in proof:} The three-loop free energy of hot
Yang-Mills theory has been obtained in Ref.\cite{AZ}. \\

Acknowledgments: \\
This project has spanned a large space-time
interval and we are happy to acknowledge the many people
who have helped one way or another, either by discussion or
hospitality : J.P. Blaizot,  E. Braaten, T.H. Hansson,
E. Iancu, L.D. McLerran, J-Y. Ollitrault, R.D. Pisarski,
D.K. Sinclair, G. Sterman, A.R. White and C. Zachos.
Also, C.C. thanks the theory groups at Lund
University and CE-Saclay, and we both thank the theory groups
at the Universities of Lecce and Stockholm, for hospitality.
Finally we are grateful to  P. Arnold and C. Zhai for helpful
discussions in pointing out some errors in our earlier calculations.

\newpage
\renewcommand{\theequation}{A.\arabic{equation}}
\setcounter{equation}{0}
\noindent
\noindent {\large\bf Appendix A}
\vskip 3mm \noindent

{ \bf A1. Identities for statistical factors \\}
Let $n_x=1/(e^x-1)$ and $N_x=1/(e^x+1)$. Also, denote the
step function by
$\theta(x)$ and the sign function by $\epsilon(x)$.  We have

\begin{eqnarray}
n_{-x}&=&-(1+n_x), \label{a1} \\
N_{-x}&=&1-N_x, \\
n_x &=&-\theta(-x)+\epsilon(x)n_{|x|},\\
N_x &=&\theta(-x)+\epsilon(x)N_{|x|},\\
n_{x+y}(n_x-n_{-y})&=& n_x n_y, \label{a5} \\
n_{x+y}(N_{-x}-N_{y})&=& N_x N_y, \\
N_{x+y}(N_{-x}+n_y) &=& N_x n_y, \label{a7} \\
n_{x+y+z}(n_{x}n_z + n_{-y} n_{-z} - n_x n_{-y}) &=& n_x
n_y n_z , \label{a8} \\
N_{x+y+z}(N_{-x}N_{-y}-N_{-x}N_z +N_y N_z)&=&N_x N_y N_z. \label{a9}
\label{ens}
\end{eqnarray}
The equations (\ref{a1})-(\ref{a7}) follow from the definitions
of $n_x$ and
$N_x$ while the last two are obtained by iterating
(\ref{a5})-(\ref{a7}).\\

{\bf A2. Standard Results \cite{GR} \\}
\begin{eqnarray}
\sum_{n=1}^{\infty}{1\over n^\alpha} &=& \zeta(\alpha)
 \, , \label{a10} \\
&&\nonumber \\
\sum_{n=1}^{\infty}{(-1)^{n+1}\over n^\alpha} &=&
\left(1-2^{1-\alpha}\right) \zeta(\alpha) \, , \label{a11} \\
&&\nonumber \\
{\cal M}_1(\alpha)  \equiv  \int_{0}^{\infty}dx
{x^\alpha\over e^x-1}  &=&
\Gamma(\alpha+1)\zeta(\alpha +1) \, ,  \label{a12} \\
&&\nonumber \\
{\cal M}_2(\alpha) \equiv  \int_{0}^{\infty}dx
{x^\alpha\over e^x +1} &=&
 (1-2^{-\alpha})\Gamma(\alpha +1)\zeta(\alpha+1) \, , \label{a13} \\
&&\nonumber \\
\int_{0}^{\infty}dx {x^{p-1}\over e^{r x }-y} &=& {1\over {y  r^p}}
\Gamma(p) \ \sum_{n=1}^{\infty}{y^n\over n^p},\nonumber \\
&&\,\,\,\,\,\,\, (p>\,\,0,\,\, r>0,\,\,  -1\,\,<\,\,y\,\,<\,\,1) \, ,
\label{a14}\\
&&\nonumber \\
\int_{0}^{\infty}dx\,\, x^{\nu-1}e^{-\mu\,\,x}&=&{1\over
\mu^{\nu}}\Gamma(\nu),\,\,\, Re(\mu,\nu)\,\,>\,\,0. \label{a15}
\end{eqnarray}
\begin{eqnarray}
\int_{0}^{1}dx\,\,x^{\mu -1}(1-x)^{\nu-1}\left( a x + b\,\, (1-x)
+ c\right)^{-(\mu + \nu)}
&=&(a+c)^{-\mu}(b+c)^{-\nu}B(\mu,\nu), \nonumber \\
&&\,\,\,\,\,\,\,\,a\,\geq 0,\,\,\,\, b\geq 0\,\,\,\, c\,>\,\,0,
\nonumber \\
&& \,\,\,\,\,\, \,\, Re\,\,\mu>\,\,0, \,\,\,Re\,\,\nu\,>\,\,0.
\nonumber \\
\label{a16}
\end{eqnarray}

{\bf A3. Derived Relations \\}
\begin{eqnarray}
{\cal M}_3(\alpha, \beta) \equiv \int_{0}^{\infty} \int_{0}^{\infty}
dx\,\,dy\,\,x^{\alpha}\,\,y^{\beta}\ n_{x+y} &=&
\Gamma(\alpha+1) \ \Gamma(\beta+1) \ \zeta(\alpha + \beta +2)
\, , \label{r1}\\
&& \nonumber \\
{\cal M}_4(\alpha, \beta) \equiv \int_{0}^{\infty} \int_{0}^{\infty}dx\,\,
dy\,\,x^{\alpha}\,\,y^{\beta}N_{x+y} &=&
\left(1-2^{-(1+\alpha+\beta)} \right)\Gamma(\alpha+1)\ \Gamma(\beta+1)
\ \zeta(\alpha +\beta +2)
\nonumber \\
&& \label{r2}
\end{eqnarray}
The result (\ref{r1}) is obtained by using in sequence the relations
(\ref{a14}), (\ref{a15}) and (\ref{a10}), and similarly for
(\ref{r2}).\\

{$ \bullet$ Some Covariant Integrals.\\}
Define
\begin{eqnarray}
F_N(+,\pm)&\equiv& \int d^DK
{\delta_+(K^2) \delta_+(Q^2) \delta_+(P^2)\over K\cdot (P+Q)\pm
P\cdot Q}
{1\over (K\cdot Q)^N} \, , \\
&& \nonumber \\
F_N(-,\pm)&\equiv& \int d^DK{\delta_+(K^2)\delta_+(Q^2)
\delta_+(P^2)\over
K\cdot (P-Q)\pm P\cdot Q} {1\over (K\cdot Q)^N}.
\label{fn-}
\end{eqnarray}
Then
\begin{eqnarray}
F_N(+,\pm)&=&\delta_+(Q^2)\delta_+(P^2)
\left({P\cdot Q\over 2}
\right)^{D/2-2 -N} \ C_N(+,\pm), \label{f1}  \\
&& \nonumber \\
F_N(-,\pm)&=&\delta_+(Q^2)\delta_+(P^2)
\left({P\cdot Q\over 2}
\right)^{D/2-2-N} \ C_N(-,\pm),  \label{f2}
\end{eqnarray}

where
\begin{eqnarray}
{ C_N(+,\pm) \over  \omega(D-1) \ 2^{D-5-N}} &=&
B(D/2-1,\ D/2-1-N)\ P_N(\pm), \\
&& \nonumber \\
{C_N(-,\pm) \over  \omega(D-1) \ 2^{D-5-N}} &=&
B(3 +N-D, \ D/2-1-N)\ P_N(\pm)  \\
&&\,\,\,\,\,\,\,\,-B(3+N-D,\ D/2-1)\ P_N(\mp) \, , \nonumber
\end{eqnarray}
and
\begin{eqnarray}
\omega(D) &=& { 2 \pi^{ D-1 \over 2} \over \Gamma\left({D-1 \over 2}
\right)} \, , \label{om} \\
&& \nonumber \\
P_N(+) &=& \pi \ \csc\left((N+3-D)\pi\right) \, ,
\label{aa29} \\
&& \nonumber \\
P_N(-) &=&\pi \ \cot\left((N+3-D)\pi\right) \label{a29} \; .
\end{eqnarray}

The result (\ref{f1}) is obtained as follows : Since the integral is
covariant,
 it may be evaluated in any convenient frame. Choose
$\vec{p} = \vec{-q}$.
The the only nontrivial integarls are the radial integral
$ \int_{0}^{\infty} dk$ and the angular integral
$ \int_{-1}^{1} d \cos \theta$ , where $\theta$ is the angle
between $\vec{k}$
and $\vec{q}$ . These two integrals may be decoupled
by a simple change
of variables and the angular integral then evaluated using (\ref{a16})
while the radial integrals are
\begin{eqnarray}
P_N(\pm )& \equiv & \int_{0}^{\infty} dz \ { z^{D-3-N}
\over z \pm 1} \, . \label{ar}
\end{eqnarray}
For the $+$ case the integral is standard \cite{GR}
while the $-$ case
is interpreted in the principal value sense and the result is
indicated in  (\ref{a29}). Consider instead the integral
\begin{eqnarray}
\hat{P}_N(-)&\equiv& \int_{0}^{\infty}dz {z^{D-3-N} \over z-1 \mp i0^+} \\
&& \nonumber \\
&=& P_{N}(-) \pm i \pi \, .
\label{a31}
\end{eqnarray}
Compared to $P_{N}(-)$, the integral $\hat{P}_{N}(-)$ has the
original pole at
$z=1$ shifted above or below the real-axis and this shift may be viewed
as a ``regularisation'' of $P_{N}(-)$ which is then given
by the real part of
$\hat{P}_{N}(-)$. Combining eqns. (\ref{aa29}-\ref{a29})
and (\ref{a31}),
we can relate $\hat{P}_{N}(-)$ to $P_{N}(+)$ which is regular ,
\begin{equation}
\hat{P}_{N}(-)= e^{\mp i\, \pi (D-3-N)}  \ P_{N}(+) \, .
\label{a322}
\end{equation}
A direct derivation of this result can  be obtained
by an analytic continuation of $P_N(\pm)$  for $z\,<0$.
We omit this  presentation since
it requires more involved
considerations of the integrals  for $z<0$.\\

{\bf A4. Relations between Gamma and Zeta functions\\}
One has the standard formulae \cite{GR}
\begin{eqnarray}
\Gamma(-n+\epsilon)&=&{(-1)^n\over n!}\left({1\over \epsilon}
+(1 + {1\over 2}+...{1\over n} -\gamma) + O(\epsilon)\right)
\, \label{a30}
\\
\gamma &=& \lim_{z\rightarrow 1}\left(\zeta(z)-{1\over z-1}\right) =
 0.5772157... \, , \\
\pi^{1-z}\zeta(z) &=& 2^z \ \Gamma(1-z)\ \zeta(1-z)\,\,
\sin\,\,{\pi z\over 2} \, ,
\label{a32}
\end{eqnarray}
and one may deduce
\begin{eqnarray}
\zeta(0)&=&-{1\over 2},\,\,\,\,\,\,\,\, \zeta'(0)=-{1\over 2}
\ln\,\, 2\pi,\,\,\,
\,\,\,\,\,\,\,\,\,\zeta''(0)=-2.00635645..., \nonumber \\
\zeta(-1)&=&-{1\over 12},\,\,\, \zeta'(-1)=-0.16542114369...,
\nonumber\\
&&\,\,\,\,\,\,\,\,\,\,\,\,\,\,\,\,\zeta''(-1)=-0.2502044....,
\nonumber \\
\zeta(2)&=&{\pi^2\over 6},\,\,\,\,\,\,\,\zeta'(2)=-0.937548254...,
\nonumber \\
&&\,\,\,\,\,\,\,\, \,\,\,\,\,\,\,\,\,\zeta''(2)=1.9892802342...
\nonumber
\end{eqnarray}

The specific values above may be obtained by a Taylor expansion of
both sides of (\ref{a12}) and (\ref{a13}) with respect to an
appropriate value of $\alpha$, and using (\ref{a30})-(\ref{a32}).\\ \\ \\

\renewcommand{\theequation}{B.\arabic{equation}}
\setcounter{equation}{0}
\noindent
\noindent {\large\bf Appendix B}
\vskip 2mm \noindent

Here we describe how to obtain (\ref{5.7}). The contribution
of the square bracket in (\ref{split}) to (\ref{s2}) is finite as
$D\rightarrow 4$ and is given by

\begin{equation}
V \equiv \int_{-\infty}^{\infty}dx \int d^2K_\perp {1\over (-{K_\perp}^2)}
{\left[ N_{k_0} \ \theta(k_0)-N_{x q} \ \theta(x q)\right]\over
{(-K_{\perp}^{2}) \over 2 x s}+ (\sigma x + \gamma)}
\label{b1}
\end{equation}
where $s=P\cdot Q = pq - \vec{p} \cdot \vec{q} \ $ and
$ k_0={-K_\perp^2\over 2 x s}p + x q + (K_\perp)_0$.
As in the discussion of (\ref{s3}) we first decompose $K_\perp$
along two orthonormal space-like vectors. Thus we write
\begin{equation}
K_\perp=\sigma_1 e^{(1)}+\sigma_2 e^{(2)}
\end{equation}
where the $e^{i}$ are two orthonormal space-like vectors :
 $e^{(1)}\cdot e^{(2)}=0$ and $(e^{(1)})^2=(e^{(2)})^2=-1$.
However unlike the case of (\ref{s3}) the integral in (\ref{b1}) depends
explicitly on the complicated energy $k_0$ and so we now require
an explicit parametrisation of our basis in order to proceed.
Though the Lorentz symmtery of the integrals in $J_1$, of which
(\ref{b1})
is a part of, is broken by the heat bath, we still have three dimensional
rotational invariance in the $(\vec{p}$, $\vec{q})$ integrals.
Therefore we can choose the following
explicit basis  (we are grateful to Cosmas Zachos for
discussions on this point)
for the evaluation of (\ref{b1}) and its contribution
to $J_1$ :
\begin{eqnarray}
&& P_\mu=p(1,0,0,1) \nonumber \\
&& Q_\mu=q(1,0, \sin\,\,\phi, \cos\,\,\phi)\nonumber \\
&& e^{(1)}=(0,1,0,0) \nonumber \\
&& e^{(2)}=({\sin\,\,\phi\over 1-\cos\,\,\phi},0,1,
{\sin\,\,\phi\over 1-\cos\,\,\phi}).
\label{b3}
\end{eqnarray}
where $\phi$ is the angle between $\vec{p}$ and $\vec{q}$.

The basis (\ref{b3}) satisfies the requirements
\begin{equation}
P^2=Q^2=e^{(1)}\cdot P=e^{(1)}\cdot Q=e^{(2)}\cdot P=e^{(2)}\cdot Q=0 \, ,
\end{equation}
and $p_0=p, \ q_0=q \ , s=P\cdot Q=p q(1-\cos\,\,\phi) \ge 0 \;
( - \pi \le \phi \le \pi) $.
Thus (\ref{b1}) can be written as

\begin{equation}
V=\int_{-\infty}^{\infty}dx\int_{-\infty}^{\infty}d\sigma_1
\int_{-\infty}^{\infty}d\sigma_2 \ {1\over \sigma_1^2 +\sigma_2^2} \
{\left[ N_{k_0} \ \theta(k_0)-N_{x q} \ \theta(x q)\right]
\over {(\sigma_1^2 +\sigma_2^2)\over 2 x s}+ (\sigma x +\gamma)}
\end{equation}
with
$k_0={(\sigma_1^2 +\sigma_2^2)\over 2 x s}p +x q +\sigma_2 e_0^{(2)}$.

Now it is convenient to transform
from the Cartesian $(\sigma_1,\sigma_2)$ coordinates to the
polar coordinates defined by
$$\sigma_1=y \ \sin\theta \;\;\;\;\; , \,  \sigma_2=y \ \cos\theta $$
and then change variables $y^2\rightarrow y$ to obtain
\begin{equation}
V=\int_{-\infty}^{\infty}dx\int_{0}^{\infty}{ dy\over y}
\int_{0}^{\pi}d\theta \ {\left[ N_{k_0} \ \theta(k'_0)-N_{x q} \
\theta(x q)\right]
\over { y \over 2 x s}+ (\sigma x +\gamma)}
\label{b5}
\end{equation}
with
$k'_0={y p\over 2 x s}+x q +\sqrt{y} \cos \ \theta \, e_0^{(2)}$.

Next split the $x-$integral in (\ref{b5}) as
$\int_{0}^{\infty}dx + \int_{-\infty}^0 dx $ and in the
second part do $x\rightarrow -x$, $\theta \rightarrow\pi -\theta$,
and note that eventually we need $\sum_{\gamma = \pm 1}V$,
to get (after some
simplification)
\begin{equation}
\sum_{\gamma= \pm 1} V =  \sum_{\gamma= \pm 1} \ \int_{0}^{\infty}dx
\int_{0}^{\infty}{dz\over z}
\int_{0}^{\pi}d\theta \ {\left[ \epsilon(\hat{k}_0) \ N_{|\hat{k}_0|}-N_{x q}
\right] \over z +(\sigma x +\gamma)}
\label{b6}
\end{equation}

where
\begin{eqnarray}
\hat{k}_0 &=& p z + x q +\sqrt{2 s x z } \ \cos\,\theta \ e_0^{(2)}\nonumber \\
&& =p z + x q +\sqrt{2 p q x z (1 + \cos\,\,\phi)}\ \cos\,\,\theta.
\label{b7}
\end{eqnarray}

Furthermore it follows from (\ref{b7}) that $\hat{k}_0 \geq\,\,0$, and thus
we can write (\ref{b6}) as
\begin{equation}
\sum_{\gamma=\pm 1} V = \sum_{\gamma = \pm 1} \
\int_{0}^{\infty}dx \int_{0}^{\infty}{dz\over z}
\int_{0}^{\pi}d\theta \ {\left( N_{\hat{k}_0}-N_{x q}\right)\over
z +(\sigma x + \gamma)} \ .
\label{b8}
\end{equation}

Finally $J_{1A}$ is given by
\begin{eqnarray}
J_{1A} &=&\int {d^3p \ d^3 q \over 4 p q} \ {\left(N_p +n_p\right)n_q
\over (2 \pi)^9} \sum_{\gamma=\pm 1}V \nonumber \\
& &={ (4\pi)(2\pi)\over 4 (2 \pi)^9}\int_{0}^{\infty}dp \ p \ (N_p + n_p)
\int_{0}^{\infty} dq \ q \ n_q\int_{-1}^{1} d\, \cos\phi
\sum_{\gamma=\pm 1}V \, ,
\label{b9}
\end{eqnarray}
with $V$ defined through equations (\ref{b8}) and (\ref{b7}).
The integrals in (\ref{b9}) can now be performed numerically.
One technical point that should be noted is the principal value singularity
in the factor ${1\over z + (\sigma x +\gamma)}$ occurring in
(\ref{b8}) and (\ref{b9}). Consider, for example,
the case $P{1\over z-(x-1)}$ in (\ref{b8}).
Doing the change of variables $z=y |x-1|$ gives us the factor
$P{1\over y-1}$ to deal with in the new $\int_{0}^{\infty}dy$ integral.
We use the definition of the principal value to write
\begin{eqnarray}
&& \int_{0}^{\infty}dy \ P{1\over y-1}\times \mbox{(rest)} \nonumber \\
&&=\lim_{\delta\rightarrow 0}\left(\int_{0}^{1-\delta}
{dy\over y-1} \mbox{(rest)} + \int_{1+\delta}^{\infty}
{ dy\over y-1} \mbox{(rest)} \right) \, ,
\label{b10}
\end{eqnarray}
where ``(rest)'' denotes the rest of the integrand in (\ref{b8}) after
the change $z=y|x-1|$.
In the second term in (\ref{b10}) one can do the change of
variables $y\rightarrow {1\over y}$ and then combine the result
with the first term to get an integral of the form

\begin{equation}
\int_{0}^{1}{dy\over y-1} \mbox{(new rest)} \, ,
\label{b11}
\end{equation}
where the limit $\delta\rightarrow 0$
can be taken because now the integral is finite as $y\rightarrow 1$ since
$\mbox{(new rest)} \rightarrow(y-1)$ as $y\rightarrow1$.
Once the principal value singularity has been removed by transforming
(\ref{b10}) into (\ref{b11}), the numerical integration of (\ref{b9})
can proceed and we obtain
\begin{equation}
J_{1A}={6.101\over 2^7 \pi^6}.
\end{equation}

Some remarks are in order. The change of variables
$z=y|x-1|$ that yielded (\ref{b10}) from (\ref{b9})
results in an
$x$-integral of the form $ \int_{0}^{\infty} dx (\cdot)/ |x-1| $. In this case
one appears to have created a new singularity at $x=1$. However this is not
true because the
rest of the integrand actually compensates it as $x \rightarrow 1$.
In other places in this paper, for example in
the evaluation of $J_{1B}$ (\ref{cstar})
and also $F_N(-,\pm)$ (\ref{fn-}) { \it near} $D=4$ dimensions,
the change $z=y |x-1|$ actually creates
a singularity at $x=1$ when  $D=4$. Noting that
$ \int_{0}^{\infty} dx (\cdot)/ |x-1| $ actually means
$ \int_{0}^{1} dx (\cdot)/(1-x) + \int_{1}^{\infty} dx (\cdot)/(x-1)$,
we have used dimensional continuation separately for each part
to regulate the singularity at $x=1$ in those cases.

For another example, consider an integral of the form
\begin{equation}
 \int_{0}^{\infty} {dz \over z^2} \int_{0}^{\infty} dx
\left[ {1 \over z-(x-1)} + {1 \over z +(x-1)} \right] ( \star) \, ,
\label{barell}
\end{equation}
which appears in the evaluation of $J_2$ (\ref{s5}). If one now
naively does the change $z=y|x-1|$ to simplify the integrals, the
$x$-integral becomes $ \int_{0}^{\infty} {dx \over |x-1|^2 } ( \mbox{new}
\star)
$
and the singularity at $x=1$ is { \it not} compensated. The safe way to proceed
in this case is to first do $x \rightarrow xz$ in (\ref{barell}) to get
\begin{equation}
\int_{0}^{\infty} dx  \int_{0}^{\infty} { dz \over z} \left[ {1 \over
 z(1-x)+1)} + {1 \over z(x+1) -1)} \right] ( \star : x \rightarrow xz ) \, .
\label{barellz}
\end{equation}
Then do $z \rightarrow z|x-1|$ for the first term, $z \rightarrow
z(x+1)$ for the second, and finally proceed as per (\ref{b10})-(\ref{b11})
to remove the principal value prescription so that the integrals may be
handled  numerically.

\pagebreak
\noindent{\underline{\Large Figure Captions}}
\smallskip

\underline{Fig.1}:\\ Contribution to order $e^2$.
The wavy line represents the photon propagator.

\underline{Fig.2}:\\
The order $e^4 N$ contributions $G_1$ and $G_2$.

\underline{Fig.3}: \\
Ultraviolet counterterm diagrams for Fig.2.

\underline{Fig.4}:\\ Fig.4a is the $e^4 N^2$ contribution
($G_3$) while Fig.4b is the corresponding counterterm
diagram $X_1$.

\underline{Fig.5}: \\
Diagrams obtained by self-energy insertions along one photon
propagator in the diagrams of Fig.2.

\vfil

\pagebreak

\end {document}